\documentclass[11pt,showpacs]{revtex4-1}

\usepackage{epsfig}
\input epsf.tex
\usepackage{amssymb}
\usepackage{amsmath}
\usepackage{amsthm}
\usepackage{amsopn}

\def\comment#1{}

\def\beq{\begin{equation}}
\def\eeq{\end{equation}}
\def\bea{\begin{eqnarray}}
\def\eea{\end{eqnarray}}

\usepackage{ulem}
\usepackage{cancel}
\usepackage{color}

\begin{document}

\title{  Relativistic Generalization of the Born Rule }
\author{  M. J. Kazemi $^{1}$\footnote{kazemi.j.m@gmail.com}, M. H. Barati$^{2}$\footnote{mohbarati14@gmail.com},  Y.Rokni$^{3}$\footnote{s.y.rokni@gmail.com } and Jafar Khodagholizadeh $^{4}$\footnote{ Lecturer in Farhangian University, Tehran,Iran. Email: gholizadeh@ipm.ir}}
\affiliation{$^{1}$ Department of Physics, Shahid Beheshti University, G. C., Evin,Tehran 19839, Iran.\\
$^{2}$
Department of Physics, kharazmi University, Tehran, Iran.\\
$^{3}$School of physics, Semnan University , Tehran, Iran.\\
$^{4}$School of physics, Institute for research in fundamental sciences (IPM), Tehran, Iran}

\begin{abstract}
We have shown that the Born rule $(\rho=|\psi|^{2})$ is inconsistent with Lorentz symmetry of the Salpeter equation (square root Klein-Gordon equation). So we find relativistic modification of the Born rule as  $\rho=\frac{1}{2mc^{2}}(|\sqrt{\widehat{H}-mc^{2}}\psi|^{2}+|\sqrt{\widehat{H}+mc^{2}}\psi|^{2})$ , which is consistent with Lorentz symmetry of this equation.
\end{abstract}


\maketitle

\section{Introduction}
The simplest relativistic generalization of the Schr\"{o}dinger equation is the "square root Klein-Gordon equation":
\begin{eqnarray}
i\hbar\dfrac{\partial \phi (\textit{\textbf{p}},t)}{\partial t}=H(\textit{\textbf{p}})\phi(\textit{\textbf{p}},t)
\end{eqnarray}
In which $ \phi (\textit{\textbf{p}},t)$ is the wave function in momentum space representation and $H(\textit{\textbf{p}})$ is the relativistic Hamiltonian;  $ H(\textit{\textbf{p}})=\sqrt{c^{2}\textit{\textbf{p}}^{2}+m_{0}^{2}c^{4}}$ . This equation can be rewritten in position space representation as follow:
 \begin{eqnarray}\label{2}
i\hbar\dfrac{\partial \psi  (\textbf{x},t)}{\partial t}=mc^{2}\sum _{k=0}^{\infty} (\dfrac{i\hbar}{m c})^{2k}\begin{pmatrix}
  1/2 \\
  k
\end{pmatrix} \boldsymbol{\nabla }^{2k}\psi  (\textbf{x},t)
\end{eqnarray}
The above equation, sometimes called "Salpeter" equation. This equation and its integral representations can be uniquely derived from following simple assumptions: (1) the equation has the first order of time.
  (2) The equation is linear. (3) The equation leads to relativistic dispersion relation for plane waves (in other words, being consistent with de Broglie relations).
The history of the square root Klein-Gordon equation (SRKG equation) goes back to the early years of Relativistic quantum mechanics. In the 1927 Weyl proposed the operator $\sqrt{-c^{2\hbar^{2}}\nabla^{2}+m^{2}c^{4}}$  for the formulation of relativistic quantum mechanics \cite{1}. But he did not expand his idea as a comprehensive theory and also other pioneers of quantum mechanics used different methods for the formulation relativistic quantum mechanics which led to other wave equations such as Dirac and Klein-Gordon equations. On the other hand, although the SRKG equation lacks proper theoretical characteristics and even gets rid of some difficulties that other relativistic wave equations have (such as Klein paradox and "Zitterbewegung"), but higher order derivatives in this equation makes it look more complicated than the other quantum wave equations; therefore it attracted less attention during the historical development of Relativistic quantum mechanics. But after derivation of this equation from Bethe-Salpeter formalism \cite{2,3,4}, it was more attention. In this regard, in the recent years, theoretical characteristics and integral representations of this equation have been attracting special attention \cite{5,6,7,8,9,10}. Also, this equation has been successfully used in description of several relativistic problems and phenomena; such as relativistic harmonic oscillator \cite{11,12,13}, waves in relativistic quantum plasma \cite{14}, relativistic (quark) bound states \cite{15,16,17,18,19,20,21}, and the relativistic Bohmian mechanics \cite{22}. We particularly note the consistency of the results of this equation with the experimental spectrum of mesonic atoms \cite{15}. In addition, since this equation is of first order with respect to time, there is the possibility of using the Born interpretation, $(\rho=|\psi|^{2})$, for the interpretation of the wave function and as a result it avoid  of negative probability problem of the Klein-Gordon equation. In this regard, K. Kowalski and J. Rembielinski derived, the position Born's probability density and its corresponding current density in the momentum representation as follow \cite{7}:

\begin{eqnarray}\label{3}
\rho_{B}=\dfrac{1}{2\pi \hbar}\int_{-\infty}^{\infty}\int_{-\infty}^{\infty}\phi^{*}(\textbf{p}_{1})\phi(\textbf{p}_{2})e^{\dfrac{i}{\hbar}(\textbf{p}_{1}-\textbf{p}_{2}).\textbf{x}}d\textbf{p}_{1}d\textbf{p}_{2}
\end{eqnarray}
\begin{eqnarray}\label{4}
\textit{\textbf{J}}_{B}=\dfrac{1}{2\pi \hbar}\int_{-\infty}^{\infty}\int_{-\infty}^{\infty}\textbf{u}(\textbf{p}_{1},\textbf{p}_{2})\phi^{*}(\textbf{p}_{1})\phi(\textbf{p}_{2})e^{\dfrac{i}{\hbar}(\textbf{p}_{1}-\textbf{p}_{2}).\textbf{x}}d\textbf{p}_{1}d\textbf{p}_{2}
\end{eqnarray}
In which the "velocity" ,$  \textit{\textbf{u}}(\textit{\textbf{p}}_{1}+\textit{\textbf{p}}_{2})$ defined as                   $ \textit{\textbf{u}}=((\textit{\textbf{p}}_2+\textit{\textbf{p}}_1 ) c^2)/(H(\textit{\textbf{p}}_2 )+H(\textit{\textbf{p}}_1 ) )$. This very possibility of a probabilistic interpretation of the wave function has been often the stimulus to using this equation.
On the other hand, the Lorentz invariance of the SRKG equation has always been under discussion Because of its high derivatives, in fact checking Lorentz invariance of this equation is complicated and cannot be specified easily. Therefor sometimes because of the inequality of time and space derivatives, this equation has been mistakenly accepted as a frame dependent   equation and incompatible with special relativity \cite{19,23,24,25,26,27,28}. But it has been shown that if the wave function is scalar this equation is Lorentz invariant (in the absence of interactions) \cite{29,30}. But we need to pay attention that Born rule, $\rho_B= |\psi|^2$, is inconsistent with assuming the wave function as scalar. Because the position probability density is the first component of the probability current four-vector and cannot be scalar. On the other word, because of the relativistic length contraction, the probability density cannot be scalar. Hence, Born rule (eq.(\ref{3}) and eq.(\ref{4})) cannot be a proper interpretation for the scalar wave function. In fact in the rest of this article we will show even if the wave function is not scalar, Born rule will still not be proper for the probabilistic interpretation of the wave function. It means that no transformation will be found for the wave function in a way to cause Born probability density and Born current densities(eq.(\ref{3}) and eq.(\ref{4})) become altogether as one four-vector. So, the Born rule must be modified to conform with Lorentz symmetry. In this regard, we find a proper relativistic modification of the Born rule for scalar wave function (spin 0 particles) in the section (I) and for spin 1/2 particles in the section (II).

\section{. The relativistic generalization of the Born rule for scalar wave function.}
We propose the relativistic generalization of the Born rule as follows:
 \begin{eqnarray}\label{5}
J^{\mu}=\dfrac{1}{2\pi \hbar}\int_{-\infty}^{\infty}\int_{-\infty}^{\infty}\textit{u}^{\mu}(\textit{\textbf{p}}_{1},\textit{\textbf{p}}_{2})\phi^{*}(\textit{\textbf{p}}_{1})\phi(\textit{\textbf{p}}_{2})e^{\dfrac{i}{\hbar}(\textit{\textbf{p}}_{1}-\textit{\textbf{p}}_{2}).\textit{\textbf{x}}}d\textit{\textbf{p}}_{1}d\textit{\textbf{p}}_{2}
\end{eqnarray}
in which $ \textit{\textit{u}}^{\mu}(\textit{\textbf{p}}_{1},\textit{\textbf{p}}_{2}) $,  is defined as:
\begin{eqnarray}\label{6}
u^{\mu}(\textit{\textbf{p}}_{1},\textit{\textbf{p}}_{2})=\gamma(\textit{\textbf{u}})(1,\textit{\textbf{u}})~~~~;~~~~\gamma(\textit{\textbf{u}})\equiv(1-\dfrac{\textit{\textbf{u}}^{2}}{c^{2}})^{-1/2}
\end{eqnarray}
The expression (\ref{5}) is the simplest covariant generalization of the eq. (\ref{3}) and eq.(\ref{4}). In fact by means of direct (and also long) calculation, we can see that $u^{\mu}$ is actually a four-vector and as a result,$J^{\mu}$  is a four-vector as well - of course if the wave function be scalar(see section II). It can also be simply shown that $J^{\mu}$ (as was defined above) satisfies the continuity equation. It is obvious that the first component of the $J^{\mu}$Specify the relativistic generalization of the Born rule in the momentum representation:
 \begin{eqnarray}\label{7}
\rho(x)=\dfrac{1}{2\pi \hbar}\int_{-\infty}^{\infty}\int_{-\infty}^{\infty}\gamma(\textit{\textbf{p}}_{1},\textit{\textbf{p}}_{2})\phi^{*}(\textit{\textbf{p}}_{1})\phi(\textit{\textbf{p}}_{2)}e^{\dfrac{i}{\hbar}(\textit{\textbf{p}}_{1}-\textit{\textbf{p}}_{2}).\textit{\textbf{x}}}d\textit{\textbf{p}}_{1}d\textit{\textbf{p}}_{2}
\end{eqnarray}
The question that arises at this point is whether the definition (\ref{7}) lead to a positive definite probability density. For answering this question, we note that the coefficient $\gamma$ in the expression (\ref{7}) can be separated as follow:
\begin{eqnarray}\label{8}
\gamma(\textit{\textbf{p}}_{1},\textit{\textbf{p}}_{2})=D^{+}(\textit{\textbf{p}}_{1})D^{+}(\textit{\textbf{p}}_{2})+D^{-}(\textit{\textbf{p}}_{1})D^{-}(\textit{\textbf{p}}_{2})
\end{eqnarray}
Where  $D^{\pm} (\textit{\textbf{p}})=(\frac{H(\textit{\textbf{p}})\pm mc^2}{2mc^{2}} )^{1/2}$.Through the aid of this separation, the position probability density can be derived in position space representation as follow:
\begin{eqnarray}\label{9}
\rho(x)=\dfrac{1}{2}[|(\sqrt{(il_{c}\nabla)^{2}+1}+1)^{\frac{1}{2}}\psi |^{2} +|(\sqrt{(il_{c}\nabla)^{2}+1}-1)^{\frac{1}{2}}\psi|^{2}]
\end{eqnarray}
Where $l_c$ is Compton wavelength; $l_c=\hbar/mc $. So  eq.(\ref{7}) leads to a positive-definite probability density. It is clear that in non- relativistic limit, $(c\rightarrow \infty)$ ,the equation (9) reduce to Born rule, $\rho=|\psi|^{2}$ . In addition, if $|\textit{\textbf{p}}_{1}-\textit{\textbf{p}}_{2} |\ll mc$  then  $\gamma(\textit{\textbf{p}}_{1},\textit{\textbf{p}}_{2})\approx 1$ and the expression (\ref{7}) gives the non-relativistic expression (\ref{3}). It means that whenever the width of wave packet in momentum space is small in comparison with  $mc$ ,$(\Delta (p)\ll mc)$ , deviation of $\rho$ from $|\psi|^{2}$ will be negligible. As a result, based on uncertainty relations, deviation of $\rho$ from $|\psi|^{2}$ is considerable only if the width of wave packet is smaller than the Compton wavelength $(\Delta (x)\ll l_{c})$. to see this fact explicitly, we plot the relativistic (eq.(\ref{9})) and nonrelativistic (Born rule) probability density for a particle in one dimensional infinite square well potential. From fig. (1) it is obvious that if the width of the box, $L$ , is greater than the Compton wavelength of the particle, $l_c$ , the diagram of the relativistic probability density, $\rho$ , will be coincident with the diagram of the Born probability density, $\rho_B$ ; and whenever the width of the box is smaller than the Compton wavelength then $\rho$ will deviates from $\rho_B$. In fact if$\frac{L}{l_{c}}\rightarrow 0$ then the probability of the particle's presence in the box will be uniform(see figure.1).
\begin{figure}
\includegraphics[scale=0.7]{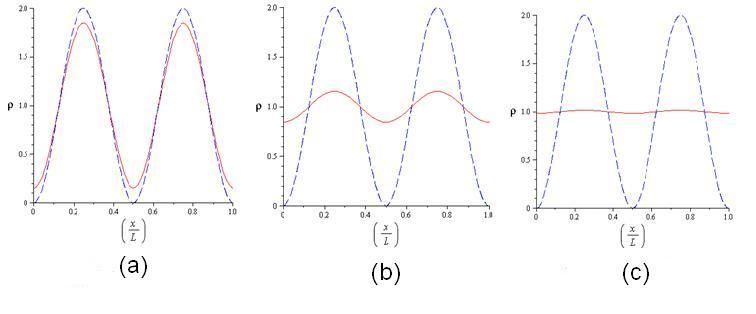}
\caption{The dashed line diagram (blue) represents non-relativistic probability density , $ \rho_{B}=|\psi|^{2}$ ; Red diagrams represent the relativistic probability density,$\rho$ , for the first excited state n=2 of particle in infinite potential well. The diagram is drawn considering different width of the box: (a)$L=10 l_{c}$,  (b)$L=l_{c}$ and (c)$L=0.1 l_{c}$.}
\label{fig:1}
\end{figure}

\section{Relativistic generalization of the Born rule for spin $ \frac{1}{2} $ particles }
In the previous section we considered the wave function as a scalar function and with this assumption we have introduced a natural generalization for the Born rule. But the wave function is not necessarily scalar. In fact if we eliminate constraints of being scalar, there are more possibilities for Relativistic generalization of the Born rule. For this purpose, without considering any transforming properties for wave function as a presumption, we consider following general form for relativistic generalization of the Born rule:
 \begin{eqnarray}\label{10}
\textit{\textbf{J}}^{\mu}=\dfrac{1}{2\pi \hbar}\int_{-\infty}^{\infty}\int_{-\infty}^{\infty} F(\textit{\textbf{p}}^{\prime},\textit{\textbf{p}}^{\prime\prime}) \gamma(1,\textit{\textbf{u}}) \phi ^{*}(\textit{\textbf{p}}^{\prime}) \phi(\textit{\textbf{p}}^{\prime \prime}) e^{\dfrac{i}{\hbar}(\textit{\textbf{p}}^{''}-\textit{\textbf{p}}^{'}).\textit{\textbf{x}}}d\textit{\textbf{p}}^{\prime}d\textit{\textbf{p}}^{\prime\prime}
\end{eqnarray}
in which $F(\textit{\textbf{p}}^{'},\textit{\textbf{p}}^{''})$ is an Unknown function that should be determined by physical constraints. It can be easily shown that the above expression satisfies the continuity equation. In fact the expression (\ref{10}) is a very simple and complete generalization of the Born rule. Now we want to see that what limitations are applied to the form of the function $F(\textit{\textbf{p}}^{'},\textit{\textbf{p}}^{''})$ by Lorentz symmetry. For this purpose, without losing the totality of the problem we will represent the wave function as the superposition of plane waves in tow inertial reference frame $S$ and $S'$:
\begin{eqnarray}\label{11}
  \psi(\textit{\textbf{x}},t)=\sum _{i=1}^{N}A_{i} e^{\dfrac{i}{\hbar}(\textit{\textbf{p}}_{i}\textit{\textbf{x}}-E(\textit{\textbf{p}}_{i})t)}
\end{eqnarray}
\begin{eqnarray}\label{12}
  \psi^{'}(\textit{\textbf{x}}^{'},t^{'})=\sum _{i=1}^{N}A_{i}^{'} e^{\dfrac{i}{\hbar}(\textit{\textbf{p}}_{i}^{'}\textit{\textbf{x}}^{'}-E(\textit{\textbf{p}}_{i}^{'})t^{'})}
\end{eqnarray}
Where $E(\textit{\textbf{p}})=\sqrt{c^2 \textit{\textbf{p}}^{2}+m^{2} c^{4} }$. this choice for time evaluation of the wave functions ensure the establishment of square root Klein-Gordon equation in both frameworks. Now, by calculating $j^{\mu}$ based on the expression (10) for the above wave functions we will see that $j^{\mu}$ is only a four-vector if:
 \begin{eqnarray}\label{13}
 [\dfrac{(F_{ij}^{'})^{2}}{F_{ii}^{'}F_{jj}^{'}}][\dfrac{F_{ii}F_{jj}}{(F_{ij})^{2}}]= [\dfrac{(\gamma_{ij}^{'})^{2}}{\gamma_{ii}^{'}\gamma_{jj}^{'}}][\dfrac{\gamma_{ii}\gamma_{jj}}{(\gamma_{ij})^{2}}]
 \end{eqnarray}
And in the same time, the wave function transform as follow:
 \begin{eqnarray}\label{14}
| A_{i} ^{'} |^{2}= \dfrac{F_{ii}}{F_{ii}^{'}}\dfrac{\gamma _{ii}^{'}}{\gamma _{ii}}|A_{i}|^{2}
 \end{eqnarray}
in which $F_{ij}$ ,$F^{'}_{ij}$,$\textit{\textbf{u}}_{ij}$ and $\gamma_{ij}$ are defined as:
\begin{eqnarray}\nonumber
F_{ij}=F(\textit{\textbf{p}}_{i},\textit{\textbf{p}}_{j})~~~~,~~~~F_{ij}^{'}=F(\textit{\textbf{p}}_{i}^{'},\textit{\textbf{p}}_{j}^{'})~~~~,~~~~\textit{\textbf{u}}_{ij}=\textit{\textbf{u}}(\textit{\textbf{p}}_{i},\textit{\textbf{p}}_{j})~~~~,~~~~\gamma_{ij}=(1-(\dfrac{\textit{\textbf{u}}_{ij}}{c})^{2})^{-1/2}
\end{eqnarray}
The expression (\ref{13}) is a constraint that should satisfy the function F for the expression (\ref{10}) to be a four-vector. Choosing $F\equiv1$ (which is equivalent with the Born rule) does not satisfy the equation (\ref{13}). In fact by substituting $F\equiv1$ the equation (\ref{13}) will lead to this incorrect equation:	
\begin{eqnarray}\label{15}
 [\dfrac{(\gamma_{ij}^{'})^{2}}{\gamma_{ii}^{'}\gamma_{jj}^{'}}][\dfrac{\gamma_{ii}\gamma_{jj}}{(\gamma_{ij})^{2}}]=1~~~~~,~~~~~i\neq j
\end{eqnarray}
The incorrectness of the above equation can be numerically examined. Therefore no transformations can be found for the wave function which could cause the four Born currents become transformed into four-vectors (and simultaneously preserving the Lorentz invariance of the SRKG equation). As a result the Born rule must necessarily be modified at a relativistic level, even if the wave function not be scalar. In fact the eq. (\ref{13}) is a functional equation for the unknown function $\textit{\textbf{F}}$ and its solutions will lead us to the relativistic modification of the Born rule. The simplest solution of this equation is $\gamma(\textit{\textbf{p}}_{1},\textit{\textbf{p}}_{2})$ which was examined in the previous section and based on the eq. (\ref{14}), this solution leads to the scalar wave function. In fact the general solution of equation (\ref{13}) is very extensive and examining all its instances is beyond this article. Therefore in this article we will focus on the following interest subset of the solutions of this equation:
\begin{eqnarray}\label{16}
F_{\frac{n}{2}}(\textit{\textbf{p}}_{i},\textit{\textbf{p}}_{j})=\frac{\gamma (\textit{\textbf{p}}_{i},\textit{\textbf{p}}_{j})}{[(\gamma (\textit{\textbf{p}}_{i},\textit{\textbf{p}}_{j})-1)(\gamma (\textit{\textbf{p}}_{i},\textit{\textbf{p}}_{j})-1)]^{\frac{n}{2}}}
\end{eqnarray}
In fact the above solutions have proper theoretical characteristic for generalization of the Born rule. For example, all of them leads to positive definite probability density. It is clear that $F_{0}$ is same with$\gamma(\textit{\textbf{p}}_{1},\textit{\textbf{p}}_{2})$  and leads to scalar wave function and therefore is proper for description spin 0 particles. Now we will show that the  $F_{1/2}$ is proper for description spin 1/2 particles. For this purpose, we simplify,  $F_{1/2}$ , by using equation (\ref{8}) as follow:
\begin{eqnarray}\label{17}
F_{1/2}(p_{i},p_{j})=1+\textit{\textbf{D}}(\textit{\textbf{p}}_{i})\textit{\textbf{D}}(\textit{\textbf{p}}_{j})
\end{eqnarray}
In which $\textit{\textbf{D}}(\textit{\textbf{p}})$ is defined as $\textit{\textbf{D}}(\textit{\textbf{p}})=\frac{\textit{\textbf{p}}c}{mc^{2}+E(\textit{\textbf{P}})}$ . So, the correspondent probability and current densities of $F_{1/2}$ will be:
\begin{eqnarray}\label{18}
\rho_{1/2}&=&\psi^{*}\psi +(\hat{\textit{\textbf{D}}}\psi)^{*}(\hat{\textit{\textbf{D}}}\psi)
\end{eqnarray}
\begin{eqnarray}\label{19}
\textit{\textbf{J}}_{1/2}&=& \psi^{*}(\hat{\textit{\textbf{D}}}\psi)+\psi(\hat{\textit{\textbf{D}}}\psi)^{*}
\end{eqnarray}
Where $\hat{\textit{\textbf{D}}}=\textit{\textbf{D}}(i\hbar \boldsymbol{\nabla})$. It seems that $\rho_{1/2}$ and $\textit{\textbf{J}}_{1/2}$ are appropriate for describing spin$-1/2$ particles and the index $1/2$ is for this very reason, Because by direct calculation we can see that the solutions of the SRKG equation with interpretation based on $\rho_{1/2}$ and $\textit{\textbf{J}}_{1/2}$ and based on the correspondence
\begin{eqnarray}
\psi\longrightarrow \Psi=\begin{pmatrix}
  \psi \\
 \textit{\textbf{D}} \psi
\end{pmatrix}
\end{eqnarray}
 is equivalent with the positive energy solutions of Dirac equation. It means that the Dirac current, $J_{D}^{\mu}$ , for $\Psi$ is the same as $J_{1/2}^{\mu}$ for $\psi$ :
\begin{eqnarray}
J_{D}^{\mu}[\Psi]=J_{1/2}^{\mu}[\psi]
\end{eqnarray}
This correspondence also causes the SRKG equation to lead to Dirac equation concerning time evolution of the four-component wave function$\Psi$.
\begin{eqnarray}
i\hbar\dfrac{\partial \psi}{\partial t}=\hat{H}\psi \longrightarrow i\hbar\dfrac{\partial \Psi}{\partial t}=\hat{H}_{D}\Psi
\end{eqnarray}
In which $\hat{H}$ is representative for the Hamiltonian operator in the square equation of Klein-Gordon and $\hat{H}_{D}$ for the Hamiltonian operator in Dirac equation. For comparison between $F_{0}$  and $F_{1/2}$  , the graph of $\rho_{0}$ and$\rho_{1/2}$ is plotted for a particle in the one dimensional box in Fig (2).
\begin{figure}
\includegraphics[scale=1]{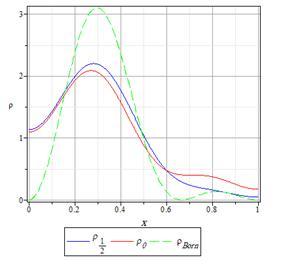}
\caption{ Compression relativistic probability density of a spin $0$ particle and spin $1/2$ particle in one dimensional box. The (dashed) green line represents Born probability density , the blue  line represent  $\rho_{1/2}$ and the Red line represent the $\rho_{0}$.  The Wave function of particle is considered as $\psi=N(\sin\frac{\pi x}{L}+\sin\frac{2\pi x}{L})$ which $L $ is the width of the box and we consider  $L=\frac{l_{c}}{2}$.}
\label{fig:1}
\end{figure}

\section{ Conclusion.}
In this article we have introduced relativistic generalization for the Born rule for particles with zero and half spins. Therefore it seems that the SRKG equation with the appropriate interpretation of the wave function can be adapted to use for particles having various spins.


\end{document}